\documentclass[aps,reprint,floatfix,twocolumn,prl]{revtex4-1}
\usepackage{xcolor} 
\usepackage{graphicx}
\usepackage[hidelinks]{hyperref}
\usepackage{amssymb}
\usepackage{amsmath}
\usepackage{ulem}

\newcommand{\be}{\begin{equation}}
\newcommand{\ee}{\end{equation}}
\newcommand{\mat}[1]{\mathrm{#1}}
\renewcommand{\vec}[1]{\mathbf{#1}}
\begin{document}
\title{Dynamically induced magnetism in KTaO$_3$}

\author{R. Matthias Geilhufe$^1$}
\author{Vladimir Juri\v ci\' c$^{1,2}$}
\author{Stefano Bonetti$^{3,4}$}
\author{Jian-Xin Zhu$^5$}
\author{Alexander V. Balatsky$^{1,6}$}
\affiliation{$^1$Nordita,  KTH Royal Institute of Technology and Stockholm University, Roslagstullsbacken 23,  10691 Stockholm,  Sweden\\
$^2$Departamento de F\'isica, Universidad T\'ecnica Federico Santa Mar\'ia, Casilla 110, Valpara\'iso, Chile\\
$^3$Department of Physics, Stockholm University, 10691 Stockholm, Sweden \\
$^4$ Department of Molecular Sciences and Nanosystems,
Ca’ Foscari University of Venice, 30172 Venice, Italy\\
$^5$ Theoretical  Division  and  Center  for  Integrated  Nanotechnologies, Los  Alamos  National  Laboratory,  Los  Alamos,  New  Mexico  87545,  USA
$^6$Department of Physics, University of Connecticut, Storrs, CT 06269, USA}

\date{\today}

\begin{abstract}
Dynamical multiferroicity features entangled dynamic orders: fluctuating electric dipoles induce magnetization. Hence, the material with paraelectric fluctuations can develop magnetic signatures if dynamically driven.   We identify the  paraelectric KTaO$_3$ (KTO) as a prime candidate for the observation of the dynamical multiferroicity. We show that when a KTO sample is exposed to a circularly polarized laser pulse, the dynamically induced ionic magnetic moments are of the order of 5\% of the nuclear magneton per unit cell. We determine the phonon spectrum using ab initio methods and  identify T$_{1u}$ as relevant soft phonon modes that couple to the external field and induce magnetic polarization. We also predict a corresponding electron effect for the dynamically induced magnetic moment which is enhanced by several orders of magnitude due to the significant mass difference between electron and ionic nucleus.
\end{abstract}
\maketitle
{\it Introduction.} Dynamical multiferroicity~\cite{juraschek2017dynamical}, the phenomenon  where the fluctuating electrical dipoles induce magnetization,  represents the dynamical counterpart of the Dzyaloshinskii-Moriya mechanism~\cite{KNB-PRL2005}. The origin of this effect lies in the duality between the electric and magnetic properties~\cite{jackson-EM}. Quite generally, the effect features entangled quantum orders. Most notably, displacive paraelectrics (PE) exhibiting a ferroelectric (FE) phase transition~\cite{khmelnitskii1973,Rowley2014,Chandra2017,Roussev2003,Edge2015,Rischau2017,Narayan2018,Arce2018}  can display  an elevated magnetic response induced by either quantum~\cite{dunnett2019dynamic} or thermal fluctuations~\cite{khaetskii2020thermal} close to the critical point. On the other hand, the dynamical magnetization can be induced by externally driving  the material, e.g. by  applying the light or a lattice strain~\cite{juraschek2017dynamical}. Dynamic multiferroicity is an example of the nonlinear phononics phenomenology~\cite{juraschek-PRL2017}, where a two phonon process induces magnetization.
From the perspective of the materials where dynamical multiferrroicity can be realized, the prime candidate to search for the effect is SrTiO$_3$ (STO), the paradigmatic quantum critical paraelectric where ferroelectricity is induced by displacive fluctuations.  It has been recently predicted that the dynamically induced magnetization both by external means and intrinsically,  close to the FE QCP in this material, may be in a measurable range~\cite{dunnett2019dynamic,khaetskii2020thermal}. 

In contrast to STO, KTaO$_3$ (KTO) is a quantum disordered  paraelectric at low temperatures with a significantly gapped transverse optical mode \cite{Rowley2014}. At zero stress, KTO retains its cubic structure down to helium temperatures \cite{lines2001principles}. The transition into a ferroelectric phase in KTO  can be induced as well, e.g., by impurities \cite{golovina2012defect, Prater1981} or strain \cite{Tyunina2010}. It is assumed that the ground state of KTO is a quantum disordered phase and significantly away from quantum critical fluctuations.   
Since KTO behaves as a regular quantum paraelectric quantum critical modes are gapped. Furthermore, on the paraelectric side of the quantum critical point, the fluctuations of the polarization are expected to be stronger and might give rise to a more dominant signal of a dynamically induced magnetic moment. So far, no prediction regarding the effect of a dynamically induced magnetization has been made for KTO, and this is precisely the aim of the current paper. 

\begin{figure}[t!]
    \centering
    \includegraphics[width=0.49\textwidth]{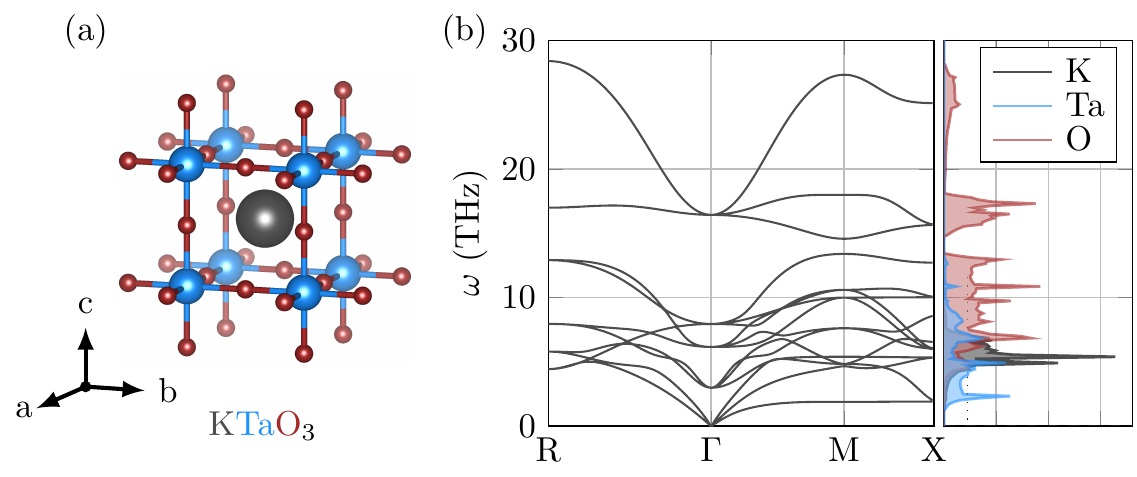}
    \caption{(a) Unit cell of KTaO$_3$. (b) Calculated phonon spectrum and phonon density of states.}
    \label{phonons}
\end{figure}

\begin{figure}[t!]
    \centering
    \includegraphics[width=0.495\textwidth]{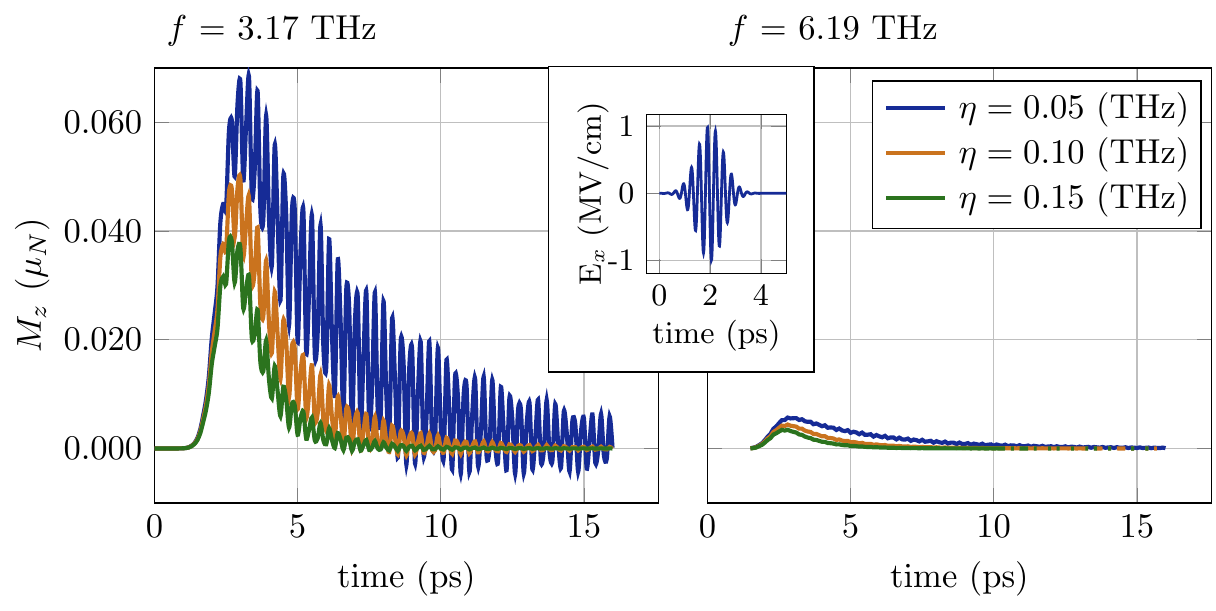}
    \caption{Dynamically induced total moment per unit cell for a laser pulse with driving frequencies 3.17 THz (left panel) and 6.19 THz (right panel).}
    \label{pulsedresult}
\end{figure}

Following the formalism of dynamical multiferroicity \cite{juraschek2017dynamical,dunnett2019dynamic,khaetskii2020thermal}, we investigate the induction of magnetic moments by applying circularly polarized terahertz radiation resonant with the phonon frequency that yield fluctuating local electric dipoles, according to
\begin{equation}
    \vec{M} = \alpha \vec{P}\times \frac{\partial}{\partial t}\vec{P} = \gamma \vec{u}\times m \frac{\partial}{\partial t}\vec{u}.
        \label{DMF}
\end{equation}
Here, $\vec{M}$ denotes the local magnetic moment, $\vec{P}$ the electric polarization, $\vec{u}$ the atomic displacement (associated with the relevant phonon mode in our analysis), $m$ the particle mass, while $\alpha$ and the gyromagnetic ratio $\gamma$ are the respective coupling constants. By performing an \textit{ab initio} analysis of the phonon spectrum (see Fig.~\ref{phonons}), we single out ${\rm T}_{1u}$ soft phonon modes as relevant for the dynamical multiferroicity. As we show, using both single mode approximation and the full dynamical matrix approach,  when the system is subjected to a resonant circularly polarized laser pulse (Fig.~\ref{pulsedresult}), one obtains a measurable magnetic signal. Taking a realistic value of the damping for the mode, we find that the induced magnetic moment per unit cell can reach the values of $\sim 0.05~\mu_N$, where $\mu_N$ is the nuclear magneton. We also predict an enhancement of the effect due to the coupling of the ion dynamics with the electronic one, which should be detectable experimentally.

{\it Phonon spectrum: First-principles calculation.} KTO crystallizes in a cubic lattice with space group Pm$\overline{3}$m (Fig \ref{phonons}(a)). We chose the experimental lattice constants as determined by Zhurova \textit{et al.} \cite{zhurova2000electron}, with a unit cell volume of 63.44 \AA$^3$. The phonon spectrum was calculated using Phonopy \cite{togo2015first}. The related force matrix was obtained from a $2\times2\times2$ supercell with automatically generated displacements, where forces were calculated using the Vienna ab initio simulation package VASP \cite{vasp}. The exchange correlation functional was approximated by the PBE functional \cite{perdew1996generalized}. We chose $8\times8\times8$ points for the Brillouin zone integration which corresponds to a $\vec{k}$-mesh density of $\approx 1050\,\vec{k}\text{-points}/\AA^{-3}$. We used a cut-off energy of $700$ eV. Additionally, we calculated the Hessian matrix for the energy landscape using density functional perturbation theory. This approach also provides a force matrix and phonon frequencies at the $\Gamma$ point, which we used to estimate the dynamically induced magnetization, as explained below. 

 The KTO unit cell contains 5 inequivalent sites, resulting in 15 phonon modes. We studied the symmetry of the phonon modes using GTPack \cite{gtpack1,gtpack2}. Constructing a five-dimensional permutation representation $\Gamma_p$ for the point group O$_h$ and the 5 unit cell sites and computing the direct product with the vector representation $\Gamma_v = T_{1u}$ we obtain $\Gamma_p \otimes \Gamma_v \simeq 4 T_{1u} \oplus T_{2u}$  corresponding to the expected modes at the $\Gamma$-point in the Brillouin zone \cite{gtpack2}. Using Phonopy we verify 4 T$_{1u}$ modes at frequencies 0.0 THz, 3.02 THz, 6.16 THz, and 16.38 THz, as well as one T$_{2u}$ mode at 7.94 THz. The former modes being soft but finite-frequency modes are instrumental for the dynamical multiferroicity, as shown below. The full phonon spectrum showing 3 acoustical and 12 optical modes is plotted in Fig. \ref{phonons}(b). The values are in good agreement with previous experiments on KTO \cite{farhi2000low}. These frequencies slightly change when calculated by using the density functional perturbation theory, giving 0.0 THz, 3.17 THz, 6.19 THz, 8.05 THz, and 16.53 THz. We notice that, in contrast to  STO, KTO does not give rise to negative energy modes in the phonon spectrum for the cubic phase, indicating the absence of a structural phase transition at low temperatures. After identifying the T$_{1u}$ soft phonon modes, we analyze the magnetic signal resulting from the exposure of the KTO system to an externally applied circularly polarized laser pulse. 
\begin{table}[b!]
    \centering
    \begin{tabular}{lccc}
\hline\hline
         & K & Ta & O \\
         \hline
        charge [e $\approx 1,602\times 10^{-19}$ C] & 0.867  & 4.954 & -1.940  \\
        mass [u $\approx 1,66\times 10^{-27}$ kg] & 39.1 & 180,95 & 16.0 \\
        \hline \hline 
    \end{tabular}
    \caption{Site parameters. Charge values according to DFT calculations performed in this study. }
    \label{parameters}
\end{table}

{\it Dynamical Multiferroicity.} 
The polarization contains an ionic and an electronic contribution and can be written as
\begin{equation}
    P_{i\alpha} = Z_{i\alpha\beta}^* u_{i\beta} + \epsilon_0\left(\epsilon_{\alpha\beta} -\delta_{\alpha\beta}\right) E_\beta.
\end{equation}
Here $u_{i\alpha}$ denotes a displacement of atom $i$ along the Cartesian coordinate $\alpha$. The Born effective charge $Z_{i\alpha\beta}^*$ describes the response of the macroscopic polarization per unit cell to the displacement of atom $i$, $Z^*_{i \alpha\beta} = \left. \Omega \frac{\partial P_\beta}{\partial u_{i\alpha}}\right|_{\vec{E}=0}$, with $\Omega$ the unit cell volume \cite{Ghosez1998}. The calculated Born effective charges for KTO are given in Tab. \ref{borneffective}. The electronic response of the polarization to the electric field is approximated in terms of the static dielectric tensor $\epsilon_{ij}$. Due to the cubic symmetry of the unit cell the dielectric tensor is diagonal and we obtain
\begin{equation}
\epsilon_{xx} = \epsilon_{yy} = \epsilon_{zz} = 5.4.%16.37.
\end{equation}
This value is sensitive to the chosen computational parameters, but consistent with other references \cite{osti_1207137}. $\epsilon_0 \approx 5.52 ~$e$^2$ keV$^{-1}$ \AA$^{-1}$ is the vacuum permittivity. 
\begin{table}[t!]
    \centering
    \begin{tabular}{lrrr}
\hline\hline
 & $Z^*_{xx}$ & $Z^*_{yy}$ & $Z^*_{zz}$ \\
\hline
 K &   1.13 &    1.13  &   1.13\\
 O &  -6.58 &   -1.69  &  -1.69\\
 O &  -1.69 &   -6.58  &  -1.69\\
 O &  -1.69 &   -1.69  &  -6.58\\
 Ta &  8.83 &    8.83  &   8.83\\
    \hline \hline 
    \end{tabular}    
    \caption{Calculated Born effective charges in units of the elementary charge $e$. }
    \label{borneffective}
\end{table}
We calculate atomic displacements $\vec{u}_i$ at the site $i$ using classical equations of motion,
\begin{equation}
    \ddot{u}_{i\alpha}(t) + \eta\,\dot{u}_{i\alpha}(t) + \sum_{j\beta} K_{i\alpha\,j\beta} u_{j\beta}(t) = \frac{Z_{i\alpha}}{m_i} E^*_\alpha(t).
    \label{EQM}
\end{equation}
Here, $Z_{i\alpha} = Z_i^0 + \sum_\beta Z^*_{i\alpha\beta} u_{i\beta}$, with $Z_i^0$ being the bare charge of the ion (see Tab. \ref{parameters}). $m_i$ the mass of atom $i$, $\eta$ is a damping factor, and $\mat{K}$ is the dynamical matrix. The electric field within the medium $\vec{E}^*$ is related to the vacuum electric field $\vec{E}$ by
\begin{equation}
    \vec{E}^* = \epsilon^{-1} \vec{E}.
\end{equation}
In experiments an additional loss in the field strength has to be taken into account due the polarization process. In our approach, the electric field induces a collective displacement of the ionic positions by coupling to the charge. Note that we do not include higher order corrections to the dielectric screening \cite{cartella2018parametric}.

We continue by discussing the size of the dynamically induced magnetic moment using a simplified analytical model. The full set of coupled differential equations is solved numerically afterwards. 
%\subsection{Single mode approximation}
We start by solving Eq.~\eqref{EQM} within a single-mode approximation, by considering one relevant mode $\omega_i = 2\pi f_i$, corresponding to one relevant site,
\begin{equation}
    \ddot{u}_{\alpha}(t) + \eta\,\dot{u}_{\alpha}(t) + \omega^2_i u_{\alpha}(t) = \frac{q}{m}\, E^*_\alpha(t).
    \label{EQMeffective}
\end{equation}
We choose circularly polarized light, i.e., $\vec{E}^*(t) = E^*_0 \left(\sin(\omega t),\cos(\omega t),0\right)$. In a coarse approximation, from \eqref{EQMeffective}, we notice that the displacement scales linearly with the applied field, $\vec{u} \approx \frac{q \vec{E}^*}{m\omega^2}$. For a harmonic displacement, we can estimate the corresponding time derivative as $\dot{\vec{u}}\approx \omega \vec{u}$. Using equation \eqref{DMF} and replacing the gyromagnetic ratio by $\gamma=\frac{q}{2m}$, we can estimate the asymptotic behavior for the dynamically induced magnetic moment by
\begin{equation}
 M_z \sim \frac{q^3 {\vec{E}^*}^2}{m^2 \omega^3}.
 \label{asymptotics}
\end{equation}
Hence, the effect increases quadratically in the field strength, but decreases with $\omega^{-3}$ in the driving frequency. The corresponding values for the charge $q$ and the mass $m$ for KTO are given in Tab. \ref{parameters}. The charges calculated using DFT are close to the chemistry picture of an ionic crystal, with integer oxydation states O$^{-2}$, K$^{+1}$, and Ta$^{+5}$.

Equation \eqref{EQMeffective} can be solved exactly. As we are solely interested in the contribution to the atomic displacement emerging due to exposure to an external laser field, we only keep the inhomogeneous part of the solution of Eq.~\eqref{EQM} that can be written as
\begin{equation}
\vec{u}(t)
= \frac{1}{\Delta_{\omega}^4+4\eta^2\omega^2}
\left( 
 \begin{array}{cc}
      \Delta_{\omega}^2 & - 2 \eta \omega  \\
      2 \eta \omega & \Delta_{\omega}^2  
 \end{array}
\right)
\frac{q}{m} \vec{E}^*,
    \label{phonon1}
\end{equation}
with $\Delta_{\omega}^2 = \omega^2_i-\omega^2$. Evaluating the polarization as $\vec{P}=\frac{q}{V} \vec{u}$, the $\omega$-dependent part of Eq.~\eqref{phonon1} can be interpreted as the susceptibility $\chi$, by transforming it into the well-known expression $\vec{P} = \chi \epsilon_0 \vec{E}$. Hence, we obtain for the magnetization
\begin{equation}
    M_z = \frac{q^3 \omega {\vec{E}^*}^2}{2 m^2 \left(\eta^2\omega^2+\Delta_\omega^4\right)}.
 \label{Mzanalytical}
\end{equation}
In the limit $\omega \gg \omega_i$, we obtain $\Delta_\omega^4\approx \omega^4$. Neglecting the damping term $\eta^2\omega^2 \ll \omega^4$ gives a similar expression to Eq.~\eqref{asymptotics}.

\begin{figure}[t!]
    \centering
    \includegraphics[width=0.4\textwidth]{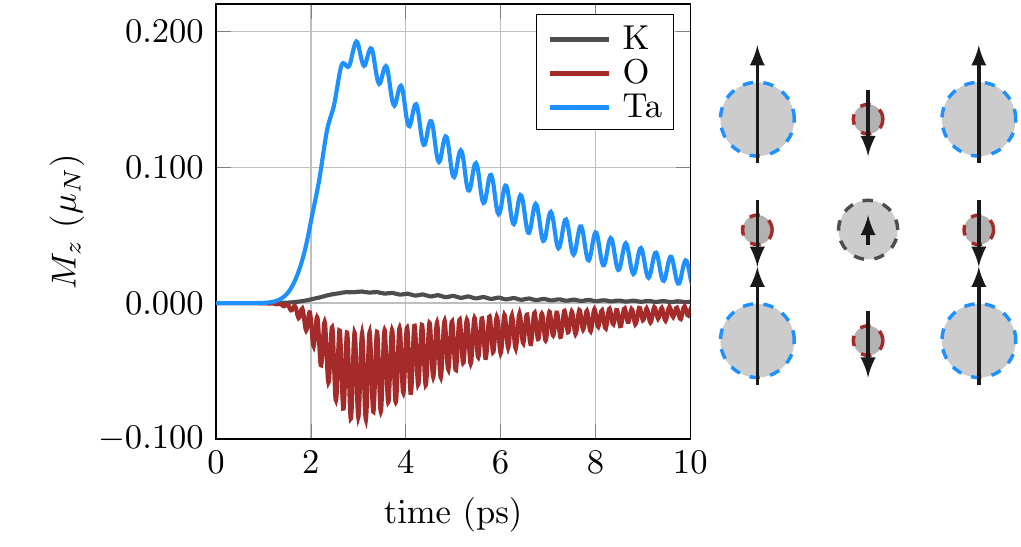}
    \caption{Site resolved dynamically induced moments within the unit cell. We used the same pulse as in Fig. \ref{pulsedresult}, a driving frequency of 3.17 THz and a damping of $\eta = 0.05$ THz.}
    \label{siteresolved}
\end{figure}

{\it System driven with a terahertz pulse.} Next, we consider a more realistic terahertz pulse and solve Eq.~\eqref{EQM} numerically. Such terahertz pulses are nowadays available \cite{SALEN20191} and allow for large peak electric field to drive phonons, but with an average deposited energy which is not enough to melt the sample. We set $\vec{u}_\alpha(0) = \dot{\vec{u}}_\alpha(0) = 0$. The pulse is modeled by a Gaussian embedding as follows,
\begin{equation}
    \vec{E}(t) = E_0\,e^{-\frac{(t-t_0)^2}{2 \sigma}} \left(
    \begin{array}{c}
         \sin(\omega t) \\ \cos(\omega t) \\ 0
    \end{array}
    \right).
\end{equation}
The considered driving frequencies are 3.17 THz and 6.19 THz, being resonant with the phonon modes. We choose a total width of $2$ ps with a peak at $2$ ps and obtain the solution for a window up to 16 ps. The dynamically induced magnetic moments are shown in Fig. \ref{pulsedresult} for various values of the damping parameter ($\eta$ = 0.05 THz, 0.10 THz, 0.15 THz). Depending on the damping factor we observe a slow decay of the dynamically induced magnetic moment. The maximal total dynamically induced magnetic moment is $\approx 0.7 \mu_N$ for small damping of $\eta < 0.1$ THz. The dynamically induced magnetic moment decreases by about one order of magnitude for a driving frequency in resonance with the $T_{1u}$ mode at 6.19 THz.
Due to the opposite local charges of the ions, the induced moments have opposite strength for O, compared to Ta and K. The site resolved dynamically induced moments due to local displacements are shown in Fig. \ref{siteresolved}. We observe that the main contributions to the total induced magnetization per unit cell come from Ta and O, being of the order of $0.2\mu_N$ and $-0.1\mu_N$ for a small value of the damping parameter, $\eta = 0.05$ THz. 

%The decreasing moment with increasing driving frequency, becomes apparent from equation \eqref{Mzmom}. Far from an eigenfrequency, 

{\it Conclusion and Outlook.} We showed that KTO is a prominent candidate for the observation of the dynamical multiferroicity. By performing an \textit{ab initio} analysis, we first find that the T$_{1u}$ soft phonon modes may be relevant for the observation of the effect. We suggest an experimental setup where the KTO sample is exposed to a circularly polarized laser field in the terahertz range to excite phonons resonantly. The dynamically induced magnetization due to locally oscillating dipoles could be measured by the time-resolved Faraday effect using a femtosecond laser pulse in the visible range. The estimated scale of the effect for an experimentally feasible setup is in the order of $10^{-2}~\mu_N$ per unit cell, with $\mu_N$ being the nuclear magneton. In Eq.~\eqref{asymptotics} we show that in an asymptotic limit, the induced moment scales quadratically with the electric field strength and to the the third power in the charge. It also scales inversely with the third power in driving frequency and the mass squared. In particular the latter feature could be of interest. 

Here we discussed the ionic movement as a driver for the induced magnetism.  We now point out an interesting possibility of induced electron motion that also would produce the magnetic moment. We expect the  angular momentum transfer from the moving ions to the electronic charge cloud in the solid. While the exact microscopic details need to be worked out the qualitative argument goes as follows.  To estimate the gyromagnetic ratio for the coupling we follow Refs. \cite{rebane1983faraday,khaetskii2020thermal} in a modified form. The position of a charged ion is denoted by $\vec{u}_+$, the average displacement of the electron cloud is $\vec{u}_-$. The respective masses are $m_+$ and $m_-$. We introduce average and relative coordinates $\vec{U}=(m_+\vec{u}_-+m_-\vec{u}_+)/(m_++m_-)$ and $\vec{u}=\vec{u}_+-\vec{u}_-$. We focus on the relative coordinate, having the momentum $p=\mu \dot{\vec{u}}$ with $\mu=m_+m_-/(m_++m_-)$. It follows for the angular momentum of the relative coordinate
\begin{equation}
    \vec{L} = \vec{u}\times\vec{p} = \frac{m_+m_-}{m_++m_-} \vec{u}\times\dot{\vec{u}}.
\end{equation}
Setting $m_-\vec{u}_+ = m_+\vec{u}_-$ we obtain for the dynamically induced moment according to Eq.~\eqref{DMF}
\begin{equation}
    \vec{M} = \vec{m}_+ + \vec{m}_- = \frac{q}{2}\frac{m_+-m_-}{m_++m_-} \vec{u}\times\dot{\vec{u}}.
\end{equation}
Taking $\vec{M} = \gamma \vec{L}$,  we obtain for the gyromagnetic ratio
\begin{equation}
    \gamma = \frac{q}{2} \left(\frac{1}{m_-} - \frac{1}{m_+}\right).
    \label{gmratio_electron}
\end{equation}
For nonequal charges, this equation generalizes to
\begin{equation}
    \gamma = \frac{m_+}{m_-} \frac{q_+}{m_++m_-} - \frac{m_-}{m_+} \frac{q_-}{m_++m_-}.
    \label{gmratio_electron2}
\end{equation}
Hence, from Eqs.~\eqref{gmratio_electron} and \eqref{gmratio_electron2} it becomes apparent that the total gyromagnetic ratio of ion and electron is dominated by the electron mass ($m_i/m_e \sim 10^3\dots 10^5$). Here we need to distinguish between a direct coupling of the electron to the external field $\sim \epsilon_0\left(\epsilon_{\alpha\beta} -\delta_{\alpha\beta}\right) E_\beta$ as well as an induced motion of the electrons due to the ionic movement. While the former contribution to the total magnetization should vanish with vanishing electric field, the latter should be present as long as the ionic movement persists. More precise analysis will be a topic of a separate publication. 

We propose KTO as a prominent candidate for the observation of the dynamical multiferroicity. Our findings open up a route for the experimental detection of the entangled dynamical orders. They should also motivate further studies of the candidate materials for the realization of the effect. 

{\it Acknowledgment.} We are grateful to G. Aeppli, U. Aschauer,   M. Basini, M. Pancaldi, O. Tjernberg,  I. Sochnikov,  N. Spaldin and J. Weissenrieder  for useful discussions.  We acknowledge support from VILLUM FONDEN via the Centre of Excellence for Dirac Materials (Grant No. 11744), the European Research Council under the European Union Seventh Framework ERS-2018-SYG
810451 HERO, the Knut and Alice Wallenberg Foundation KAW 2018.0104. 
V.J. acknowledges the support of the Swedish Research Council (VR 2019-04735) and J.-X.Z. was supported by the Los Alamos National Laboratory LDRD Program. SB acknowledges support from the Swedish Research Council (VR 2018-04611). The computational resources were provided by the Swedish National Infrastructure for Computing (SNIC) via the High Performance Computing Centre North (HPC2N) and the Uppsala Multidisciplinary Centre for Advanced Computational Science (UPPMAX).

\bibliography{references}
\end{document}